# On Stability, Persistence and Hopf Bifurcation in Fractional Order Dynamical Systems

H.A.El-Saka[1], E.Ahmed[2], M.I.Shehata[2] and A.M.A.El-Sayed[3]

1. Mathematics Department, Faculty of Science, Mansoura University, Damietta EGYPT.
2. Mathematics Department, Faculty of Science, Mansoura University, Mansoura, EGYPT.
3. Mathematics Department, Faculty of Science, Alexandria University, Alexandria, EGYPT.

Abstract:
  This is a preliminary study for bifurcation in fractional order dynamical systems. The local stability of some one and two dimensional functional equations is given. A spatial dependence of functional equations is introduced and the stability of its homogeneous equilibrium is studied. Persistence of some continuous time fractional order differential equations is studied. A numerical example for Hopf-type bifurcation in a fractional order system is given hence we propose a modification of the conditions of Hopf bifurcation.

**1.Introduction:**
  Biology [Edelstein Keshet 2004] is a rich source for mathematical ideas. We argue that fractional equations [Stanislavsky 2000] are related to complex adaptive systems (CAS) [Boccara 2004]. Such systems include biological, economic and social systems. Since persistence [Hofbauer and Sigmund 1998] and seasonality are important concepts in biology, it is relevant to study persistence of some biologically motivated non-autonomous fractional equations. In sec.2 the relation between CAS and fractional mathematics is pointed out. In sec.3 sufficient conditions for the persistence of some biologically inspired, fractional non-autonomous equations are derived. In sec.4 local asymptotic stability of some functional equations are studied.

**Definition (1)**[Smith 2003]: A complex adaptive system consists of inhomogeneous, interacting adaptive agents.
**Definition (2):** "An emergent property of a CAS is a property of the system as a whole which does not exist at the individual elements (agents) level".
  Typical examples are the brain, the immune system, the economy, social systems, ecology, insects swarm, etc…
  Therefore to understand a complex system one has to study the system as a whole and not to decompose it into its constituents. This totalistic approach is against the standard reductionist one, which tries to decompose any system to its constituents and hopes that by understanding the elements one can understand the whole system..
  Recently [Stanislavsky 2000] it became apparent that fractional equations solve some of the above mentioned problems for the PDE approach. To see this consider the following evolution equation

$$df(t)/dt = -\lambda^2 \int_0^t k(t-t')f(t')dt' \quad (1)$$

If the system has no memory then $k(t-t') = \delta(t-t')$ and one gets $f(t) = f_0 \exp(-\lambda^2 t)$.
If the system has an ideal memory then $k(t-t') = \{1 \text{ if } t \geq t', 0 \text{ if } t < t'\}$ hence
$f \approx f_0 \cos \lambda t$. Using Laplace transform $L[f] = \int_0^\infty f(t)\exp(-st)dt$ one gets L[f]=1 if there is no memory and 1/s if there is ideal memory hence the case of non-ideal memory is expected to be given by $L[f] = 1/s^\alpha$, $0 < \alpha < 1$. In this case equation (1) becomes

$$df(t)/dt = \int_0^t (t-t')^{\alpha-1} f(t')dt'/\Gamma(\alpha) \quad (2)$$

where $\Gamma(\alpha)$ is the Gamma function. This system has the following solution
$f(t) = f_0 E_{\alpha+1}(-\lambda^2 t^{\alpha+1})$,
where $E_\alpha(z)$ is the Mittag-Leffler function given by

$$E_\alpha(z) = \sum_{k=0}^\infty z^k / \Gamma(\alpha k + 1)$$

It is direct to see that $E_1(z) = \exp(z), E_2(z) = \cos z$.

Following a similar procedure to study a random process with memory, one obtains the following fractional evolution equation

$$\partial^{\alpha+1} P(x,t)/\partial t^{\alpha+1} = \sum_n (-1)^n/n! \partial^n [K_n(x)P(x,t)]/\partial x^n, \quad 0 < \alpha < 1 \quad (3)$$

where P(x,t) is a measure of the probability to find a particle at time t at position x. We expect that (3) will be relevant to many complex adaptive systems and to systems where fractal structures are relevant since it is argued that there is a relevance between fractals and fractional differentiation [Rocco and West 1999].
For the case of fractional diffusion equation the results are

$$\partial^{\alpha+1} P(x,t)/\partial t^{\alpha+1} = D \partial^2 P(x,t)/\partial x^2, P(x,0) = \delta(x), \partial P(x,0)/\partial t = 0 \Rightarrow$$
$$P = (1/(2\sqrt{Dt^\beta}))M(|x|/\sqrt{Dt^\beta}; \beta), \quad \beta = (\alpha+1)/2 \quad (4)$$

$$M(z;\beta) = \sum_{n=0}^\infty [(-1)^n z^n / \{n! \Gamma(-\beta n + 1 - \beta)\}]$$

For the case of no memory $\alpha = 0 \Rightarrow M(z;1/2) = \exp(-z^2/4)$.
Thus fractional equations naturally represent systems with memory and fractal systems which are ubiquitous properties of many CAS. Consequently they are relevant to CAS.

## 2. The Persistence results:
Definition (3): The local stability of up to 3-dimensional fractional order continuous time dynamical systems has been studied in [Ahmed, El-Saka and El-Sayed 2006]. Using Caputo's definition of fractional order derivative

$$D^\alpha f(t) = (1/\Gamma(1-\alpha)) \int_0^t [f'(s)/(t-s)^\alpha]ds, 0 < \alpha \leq 1$$

It is direct to see that $D^\alpha f(t) > 0 \ \forall t > 0 \Leftrightarrow f'(t) > 0 \forall t > 0$. Hence it is easy to prove the following:

**Proposition (1):** If the equilibrium solution of an integer order dynamical systems is globally stable by Lyapunov function V (i.e. V(0)=0, V'(t)>0 for all t>0, V'(0)=0) then its fractional order counterpart has V as its Lyapunov function (i.e. V(0)=0, $D^\alpha V(t) > 0 \forall t > 0, V'(0) = 0$).

A dynamical system is persistent if $x(0) > 0 \Rightarrow \lim_{t \to \infty} \inf[x(t)] > 0$.

Begin with the non-autonomous logistic equation
du(t)/dt= u(t)[b(t)-a(t)u(t)], u(t)>0 for all $t \geq 0$   (5)
Assume that b(t), a(t) are bounded continuous functions for all $t \geq 0$, define
$g^* = \sup\{g(t), t_0 \leq t \leq \infty\}, \ g_* = \inf\{g(t), t_0 \leq t \leq \infty\}$   (6)
then:

**Proposition (2):** If b(t) and a(t) are bounded continuous functions and if $b_* > a^* > 0$, then the following system is persistent:
$$D^\alpha u(t) = u(t)[b(t) - a(t)u(t)], \ 1 > \alpha > 0 \quad (7)$$
Proof: The solution of (7) is
$$u(t) = [1/\Gamma(1-\alpha)]\int_0^t u(s)[b(s) - a(s)u(s)]/(t-s)^{1-\alpha} ds$$
if u(t)<1 then
$$u(t) > [1/\Gamma(1-\alpha)](b_* - a^*)\int_0^t u(s)/(t-s)^{1-\alpha} ds = E_\alpha((b_* - a^*)t^\alpha) > 0$$

This completes the proof.
Similarly one can prove the following:

**Proposition (3):** If the functions $b_i(t), a_{ij}(t)$ where i,j=1,2,…,n are bounded, continuous functions and if
$$b_{*i} > \sum_{j=1}^n a_{ij}^* > 0 \ \forall i = 1,2,...,n$$
then the following system is persistent
$$D^\alpha u_i = u_i(t)[b_i(t) - u(t)\sum_{j=1}^n a_{ij}(t)], i = 1,2,...,n, 1 > \alpha > 0 \quad (8)$$

The integer order Lotka-Volterra system corresponding to (8) has been studied in [Montes de Oca and Zeeman 1995, Zhao and Jiang 2004].

3. Hopf Bifurcation
In this section we study the system
$$D^\alpha x(t) = y(t), \ D^\alpha y(t) = -x(t) + ay(t) - by^3(t), \ \alpha \in (0,1], \ 0<a<2. \quad (9)$$
It is direct to see that the null equilibrium x=y=0 is locally asymptotically stable if $(\sqrt{4-a^2})/a > \tan(\alpha \pi/2)$.

Henceforth we set $\alpha = 2/3$ hence the equilibrium solution (0,0) is locally asymptotically stable if 0<a<1. We studied the system (9) numerically for $0 < a \leq 2$, x(0)=0.1, y(0)=0.1, b=-0.3 and we obtained cycles as shown in the figure.

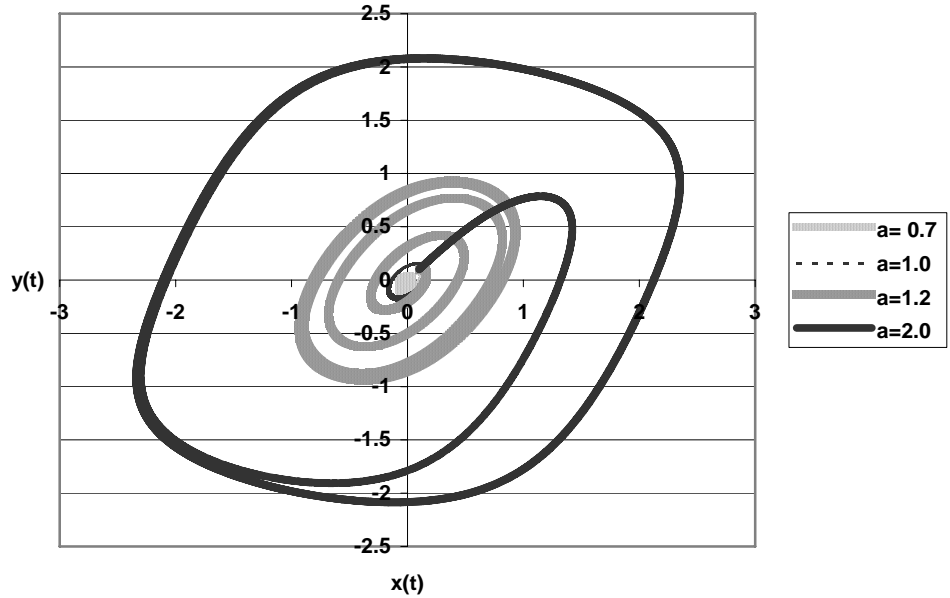

It is known that this system has a Hopf bifurcation at a=0 i.e. all the following conditions are satisfied by one of the two solutions (say $\lambda(a)$) of the characteristic polynomial of the systems (9) when linearized around (0,0):
(i) Re[$\lambda(0)$]=0, (ii) Im[$\lambda(0)$]$\neq 0$, (iii) $d\{\text{Re}[\lambda(0)]\}/da \neq 0$.

The figure shows that the cycle appeared at a=1 hence the conditions of fractional order Hopf bifurcation differs from those of the integer order case. Hence we propose the following conditions:
(i) $|\arg[\lambda(a^*)]| = \alpha\pi/2$, (ii) $d\lambda(a^*)/da \neq 0$, (iii) $|\lambda(a^*)| = 1$.

4. Local stability of functional equations:
Here we study the following functional equations:
$x(t) = f(x(t-r_1), x(t-r_2))$, $r_1, r_2 > 0$. (10)
$x(t) = f(x(t-r_1), y(t-r_2))$, $y(t) = g(x(t-r_3), y(t-r_4))$, $r_1, r_2 > 0$. (11)
Here t is a continuous variable. The equilibrium solution of (10) is given by $x_{eq} = f(x_{eq}, x_{eq})$ while the equilibrium solution of (11) is given by $x_{eq} = f(x_{eq}, y_{eq})$, $y_{eq} = g(x_{eq}, y_{eq})$.
The equilibrium solution of (10) is locally asymptotically stable if all the roots $\lambda$ of the following equation satisfy $|\lambda| < 1$ where

$$1 = (\lambda)^{-r_1} \partial f(x,y)/\partial x \big|_{x=y=x_{eq}} + (\lambda)^{-r_2} \partial f(x,y)/\partial y \big|_{x=y=x_{eq}} \quad (12)$$

The equilibrium solution of (11) is locally asymptotically stable if all the roots $\lambda$ of the following equation satisfy $|\lambda| < 1$ where

$(\partial f/\partial x \lambda^{-r_1} - 1)((\partial g/\partial y \lambda^{-r_4} - 1) - (\partial f/\partial y)(\partial g/\partial x) = 0$ (13),

where all the derivatives in (13) are calculated at the equilibrium values.
Now we give two examples:
Example (1): $x(t) = \mu x(t-r_1)(1 - x(t-r_2))$, $\mu, r_1, r_2 > 0$ (14)

The equilibrium solutions are x=0, $x = 1 - 1/\mu, \mu > 1$. The solution x=0 is locally asymptotically stable if $\mu < 1$ while the second solution is locally asymptotically stable if all the roots $\lambda$ of the following equation satisfy $|\lambda| < 1$ where

$$\lambda^{r_2} - \lambda^{r_2 - r_1} + (\mu - 1) = 0 \quad (15)$$

When $r_2 = r_1 = r$ we regain the standard condition that $x = 1 - 1/\mu, \mu > 1$ is locally asymptotically stable if $3 > \mu > 1$.

Example (2): Lotka-Volterra type functional equation:

$$x(t) = ax(t - r_1) - x(t - r_1) y(t - r_1)$$
$$y(t) = -by(t - r_2) + x(t - r_2) y(t - r_2) \quad (16)$$

Where a>1, b>0. The equilibrium solutions are (0,0) and (1+b,a-1). The null solution is unstable since a>1 while the other solution is ) is locally asymptotically stable if all the roots $\lambda$ of the following equation satisfy $|\lambda| < 1$ where

$$(\lambda^{r_1} - 1)(\lambda^{r_2} - 1) + (a - 1)(b + 1) = 0 \quad (17)$$

We propose to include spatial dependence of functional equations as follows:

$$x_i^t = (1 - D) f(x_i^{t - r_1}) + (D/2)[f(x_{i+1}^{t - r_2}) + f(x_{i-1}^{t - r_2})] \quad (18)$$

and another type of functional equation coupled map lattice may take the form:

$$x_i^t = (1 - D) f(x_i^{t - r_1}) + D \sum_{j \neq i}^{n} f(x_j^{t - r_2}) / (n - 1)] \quad (19)$$

where $D, r_1, r_2 > 0$ and i=1,2,…,n. The homogeneous equilibrium of the systems (18), (19) are given by $x_{eq} = f(x_{eq})$ and it is locally asymptotically stable if all the roots $\lambda$ of the following equation satisfy $|\lambda| < 1$ where

$$1 = f'(x_{eq})[(1 - D)\lambda^{-r_1} + D\lambda^{-r_2}] \quad (20)$$

**Proposition (4):** A sufficient condition for $x_{eq}$ to be locally asymptotically stable is $|f'(x_{eq})| < 1$.

**Proof:** If $|\lambda| \geq 1$ then $|RHS \text{ of } (20)| \leq |f'(x_{eq})| < 1$ which is a contradiction.

We like to point out that this is only a preliminary study and we hope that it will stimulate deeper studies to solve some of the following open problems in fractional order dynamical systems (FODS): Global stability in FODS (notice that the chain rule is not valid in fractional order differentiation hence straightforward generalization of Lyapunov theorem to FODS may not be possible). Can the proposed conditions for Hopf bifurcations be derived? What are the corresponding theorems to Poincare theorem for periodic orbits and Bendixon criteria for the non-existence of periodic orbits in 2-dimensions? What are the measures of chaos (in addition to sensitive dependence on initial conditions) in FODS?

**References:**

Ahmed E., A.M.A. El-Sayed, H.A.A. El-Saka,(2006)"On some Routh-Hurwitz conditions for fractional order differential equations and their applications in Lorenz, Rossler, Chua and Chen systems", Phys.Lett.A. 358,1.
Boccara N. (2004), Modeling complex systems, Springer Publ. Berlin.

Edelstein-Keshet L. (2004), "Introduction to mathematical biology",Siam Classics in Appl.Math.
Hofbauer J. and Sigmund K. (1998), "Evolutionary games and population dynamics" Cambridge Univ.Press U.K..
Matignon D. (1996) , Stability results for fractional differential equations with applications to control processing, Computational Eng. in Sys. Appl. Vol.2 Lille France 963.
Montes de Oca F. and Zeeman M.L. (1995), Balancing survival and extinction in non-autonomous competitive Lotka-Volterra systems, J.Math.Anal.Appl. 192, 360.
Rocco A. and West B.J. (1999), "Fractional Calculus and the Evolution of Fractal Phenomena"' Physica A 265, 535.
Smith J.B.(2003), "A technical report of complex system", ArXiv.CS 0303020.
Stanislavsky A.A.(2000), "Memory effects and macroscopic manifestation of randomness", Phys. Rev.E61, 4752.
Zhao J. and Jiang J. (2004), Average conditions for permanence and extinction in non-autonomous Lotka-Volterra system, J.Math.Anal.Appl. 299, 663.